\documentclass[10pt,letterpaper,twocolumn,american,nofootinbib]{revtex4}
\usepackage{ae}
\usepackage{aecompl}
\usepackage[T1]{fontenc}
\usepackage[latin1]{inputenc}
\usepackage{amsmath}
\usepackage{graphicx}
\usepackage{amssymb}

\makeatletter
\usepackage{psfrag}

\usepackage{babel}
\makeatother
\begin{document}

\title{Predictions in eternal inflation}

\author{Sergei Winitzki}

\affiliation{Department of Physics, Ludwig-Maximilians University, 80333 Munich,
Germany}

\begin{abstract}
In generic models of cosmological inflation, quantum fluctuations
strongly influence the spacetime metric and produce infinitely many
regions where the end of inflation (reheating) is delayed until arbitrarily
late times. The geometry of the resulting spacetime is highly inhomogeneous
on scales of many Hubble sizes. The recently developed string-theoretic
picture of the {}``landscape'' presents a similar structure, where
an infinite number of de Sitter, flat, and anti-de Sitter universes
are nucleated via quantum tunneling. Since observers on the Earth
have no information about their location within the eternally inflating
universe, the main question in this context is to obtain statistical
predictions for quantities observed at a random location. I describe
the problems arising within this statistical framework, such as the
need for a volume cutoff and the dependence of cutoff schemes on time
slicing and on the initial conditions. After reviewing different approaches
and mathematical techniques developed in the past two decades for
studying these issues, I discuss the existing proposals for extracting
predictions and give examples of their applications.
\end{abstract}
\maketitle

\section{Eternal inflation\label{sec:Eternal-inflation}}

The general idea of eternally inflating spacetime was first introduced
and developed in the 1980s~\cite{Vilenkin:1983xq,Starobinsky:1986fx,Linde:1986fd,Goncharov:1987ir}
in the context of slow-roll inflation. Let us begin by reviewing the
main features of eternal inflation, following these early works.

A prototypical model contains a minimally coupled scalar field $\phi$
(the {}``inflaton'') with an effective potential $V(\phi)$ that
is sufficiently flat in some range of $\phi$. When the field $\phi$
has values in this range, the spacetime is approximately de Sitter
with the Hubble rate \begin{equation}
\frac{\dot{a}}{a}=\sqrt{\frac{8\pi}{3}V(\phi)}\equiv H(\phi).\label{eq:Einstein1}\end{equation}
(We work in units where $G=c=\hbar=1$.) The value of $H$ remains
approximately constant on timescales of several Hubble times ($\Delta t\gtrsim H^{-1}$),
while the field $\phi$ follows the slow-roll trajectory $\phi_{\text{sr}}(t)$.
Quantum fluctuations of the scalar field $\phi$ in de Sitter background
grow linearly with time~\cite{Vilenkin:1982wt,Linde:1982uu,Starobinsky:1982ee},\begin{equation}
\langle\hat{\phi}^{2}(t+\Delta t)\rangle-\langle\hat{\phi}^{2}(t)\rangle=\frac{H^{3}}{4\pi^{2}}\Delta t,\label{eq:delta-phi}\end{equation}
at least for time intervals $\Delta t$ of order several $H^{-1}$.
Due to the quasi-exponential expansion of spacetime during inflation,
Fourier modes of the field $\phi$ are quickly stretched to super-Hubble
length scales. However, quantum fluctuations with super-Hubble wavelengths
cannot maintain quantum coherence and become essentially classical~\cite{Linde:1982uu,Starobinsky:1982ee,Vilenkin:1983xp,Brandenberger:1984wt,Guth:1985ya};
this issue is discussed in more detail in Sec.~\ref{sub:Physical-justifications}
below. The resulting field evolution $\phi(t)$ can be visualized~\cite{Bardeen:1983qw,Vilenkin:1983xq,Vilenkin:1983xp}
as a Brownian motion with a {}``random jump'' of typical step size
$\Delta\phi\sim H/(2\pi)$ during a time interval $\Delta t\sim H^{-1}$,
superimposed onto the deterministic slow-roll trajectory $\phi_{\text{sr}}(t)$.
A statistical description of this {}``random walk''-type evolution
$\phi(t)$ is reviewed in Sec.~\ref{sub:Eternal-inflation-from}. 

The {}``jumps'' at points separated in space by many Hubble distances
are essentially uncorrelated; this is another manifestation of the
well-known {}``no-hair'' property of de Sitter space~\cite{Gibbons:1977mu,Hawking:1981fz,Hawking:1982my}.
Thus the field $\phi$ becomes extremely inhomogeneous on large (super-horizon)
scales after many Hubble times. Moreover, in the semi-classical picture
it is assumed~\cite{Starobinsky:1986fx} that the local expansion
rate $\dot{a}/a\equiv H(\phi)$ tracks the local value of the field
$\phi(t,\mathbf{x})$ according to the Einstein equation~(\ref{eq:Einstein1}).
Here $a(t,\mathbf{x})$ is the scale factor function which varies
with $\mathbf{x}$ only on super-Hubble scales, $a(t,\mathbf{x})\Delta x\gtrsim H^{-1}$.
Hence, the spacetime metric can be visualized as having a slowly varying,
{}``locally de Sitter'' form (with spatially flat coordinates $\mathbf{x}$),\begin{equation}
g_{\mu\nu}dx^{\mu}dx^{\nu}=dt^{2}-a^{2}(t,\mathbf{x})d\mathbf{x}^{2}.\label{eq:locally de Sitter}\end{equation}

\begin{figure}
\begin{center}\psfrag{t}{$t$}\psfrag{x}{$x$}\psfrag{y}{$y$}\includegraphics[%
  width=0.70\columnwidth]{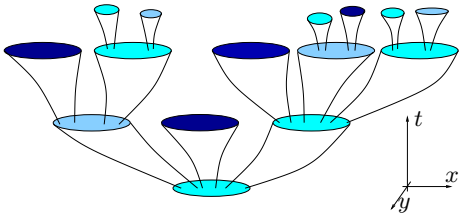}\end{center}

\caption{A qualitative diagram of self-reproduction during inflation. Shaded
spacelike domains represent Hubble-size regions with different values
of the inflaton field $\phi$. The time step is of order $H^{-1}$.
Dark-colored shades are regions undergoing reheating ($\phi=\phi_{*}$);
lighter-colored shades are regions where inflation continues. On average,
the number of inflating regions grows with time.\label{cap:A qualitative diagram} }
\end{figure}

The deterministic trajectory $\phi_{\text{sr}}(t)$ eventually reaches
a (model-dependent) value $\phi_{*}$ signifying the end of the slow-roll
inflationary regime and the beginning of the reheating epoch (thermalization).
Since the random walk process will lead the value of $\phi$ away
from $\phi=\phi_{*}$ in some regions, reheating will not begin everywhere
at the same time. Moreover, regions where $\phi$ remains in the inflationary
range will typically expand faster than regions near the end of inflation
where $V(\phi)$ becomes small. Therefore, a delay of the onset of
reheating will be rewarded by additional expansion of the proper 3-volume,
thus generating more regions that are still inflating. This feature
is called {}``self-reproduction'' of the inflationary spacetime~\cite{Linde:1986fd}.
Since each Hubble-size region evolves independently of other such
regions, one may visualize the spacetime as an ensemble of inflating
Hubble-size domains (Fig.~\ref{cap:A qualitative diagram}).

The process of self-reproduction will never result in a global reheating
if the probability of jumping away from $\phi=\phi_{*}$ and the corresponding
additional volume expansion factors are sufficiently large. The corresponding
quantitative conditions and their realization in typical models of
inflation are reviewed in Sec.~\ref{sec:Presence-of-eternal}. Under
these conditions, the process of self-reproduction of inflating regions
continues forever. At the same time, every given comoving worldline
(except for a set of measure zero; see Sec.~\ref{eternal points})
will sooner or later reach the value $\phi=\phi_{*}$ and enter the
reheating epoch. The resulting situation is known as {}``eternal
inflation''~\cite{Linde:1986fd}. More precisely, the term {}``eternal
inflation'' means future-eternal self-reproduction of inflating regions~\cite{Vilenkin:2004vx}.%
\footnote{It is worth emphasizing that the term {}``eternal inflation'' refers
to future-eternity of inflation in the sense described above, but
does not imply past-eternity. In fact, inflationary spacetimes are
generically \emph{not} past-eternal~\cite{Borde:1993xh,Borde:2001nh}.%
} To emphasize the fact that self-reproduction is due to random fluctuations
of a field, one refers to this scenario as {}``eternal inflation
of random-walk type.'' Below we use the terms {}``eternal self-reproduction''
and {}``eternal inflation'' interchangeably.

\begin{figure}
\begin{center}\includegraphics[%
  width=0.60\columnwidth,
  keepaspectratio]{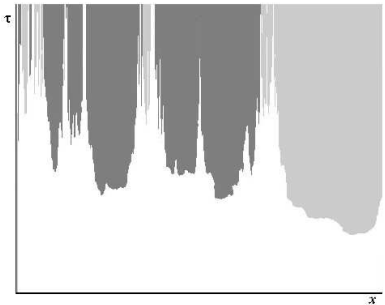}\end{center}

\caption{A 1+1-dimensional slice of the spacetime structure in an eternally
inflating universe (numerical simulation in Ref.~\cite{Vanchurin:1999iv}).
Shades of different color represent different, causally disconnected
regions where reheating took place. The reheating surface is the line
separating the white (inflating) domain and the shaded domains.\label{cap:reheating-surface-1} }
\end{figure}

Observers like us may appear only in regions where reheating already
took place. Hence, it is useful to consider the locus of all reheating
events in the entire spacetime; in the presently considered example,
it is the set of spacetime points $x$ there $\phi(x)=\phi_{*}$.
This locus is called the \emph{reheating surface} and is a noncompact,
spacelike three-dimensional hypersurface~\cite{Borde:1993xh,Vilenkin:1995yd}.
It is important to realize that a finite, initially inflating 3-volume
of space may give rise to a reheating surface having an infinite 3-volume,
and even to infinitely many causally disconnected pieces of the reheating
surface, each having an infinite 3-volume (see Fig.~\ref{cap:reheating-surface-1}
for an illustration). This feature of eternal inflation is at the
root of several technical and conceptual difficulties, as will be
discussed below.

Everywhere along the reheating surface, the reheating process is expected
to provide appropriate initial conditions for the standard {}``hot
big bang'' cosmological evolution, including nucleosynthesis and
structure formation. In other words, the reheating surface may be
visualized as the locus of the {}``hot big bang'' events in the
spacetime. It is thus natural to view the reheating surface as the
initial equal-time surface for astrophysical observations in the post-inflationary
epoch. Note that the observationally relevant range of the primordial
spectrum of density fluctuations is generated only during the last
60 $e$-foldings of inflation. Hence, the duration of the inflationary
epoch that preceded reheating is not directly measurable beyond the
last 60 $e$-foldings; the total number of $e$-foldings can vary
along the reheating surface and can be in principle arbitrarily large.%
\footnote{For instance, it was shown that holographic considerations do not
place any bounds on the total number of $e$-foldings during inflation~\cite{Lowe:2004zs}.
For recent attempts to limit the number of $e$-foldings using a different
approach, see e.g.~\cite{Wu:2006ew}. Note also that the effects
of {}``random jumps'' are negligible during the last 60 $e$-foldings
of inflation, since the produced perturbations must be of order $10^{-5}$
according to observations.%
} 

\begin{figure}
\begin{center}\psfrag{L-domination}{$\Lambda$ domination} \psfrag{reheating}{reheating} \psfrag{wall}{wall} \psfrag{nucleation}{nucleation}\includegraphics[%
  width=0.80\columnwidth]{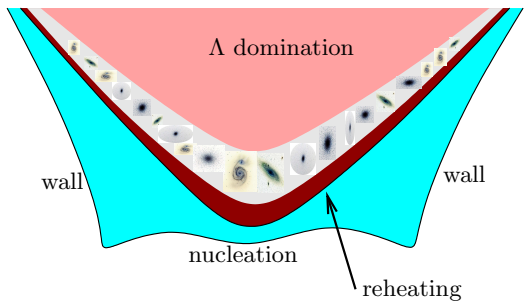}\end{center}

\caption{A spacetime diagram of a bubble interior. The infinite, spacelike
reheating surface is shown in darker shade. Galaxy formation is possible
within the spacetime region indicated. \label{cap:A-spacetime-diagram-bubble}}
\end{figure}

The phenomenon of eternal inflation is also found in multi-field models
of inflation~\cite{Linde:1991km,Linde:1993cn}, as well as in scenarios
based on Brans-Dicke theory~\cite{Garcia-Bellido:1993wn,Garcia-Bellido:1994vz,Garcia-Bellido:1995kc},
topological inflation~\cite{Linde:1994hy,Vilenkin:1994pv}, braneworld
inflation~\cite{Kunze:2003vp}, {}``recycling universe''~\cite{Garriga:1997ef},
and the string theory landscape~\cite{Susskind:2003kw}. In some
of these models, quantum tunneling processes may generate {}``bubbles''
of a different phase of the vacuum (see Sec.~\ref{sub:Bubble-nucleation-models}
for more details). Bubbles will be created randomly at various places
and times, with a fixed rate per unit 4-volume. In the interior of
some bubbles, additional inflation may take place, followed by a new
reheating surface. The interior structure of such bubbles is sketched
in Fig.~\ref{cap:A-spacetime-diagram-bubble}. The nucleation event
and the formation of bubble walls is followed by a period of additional
inflation, which terminates by reheating. Standard cosmological evolution
and structure formation eventually give way to a $\Lambda$-dominated
universe. Infinitely many galaxies and possible civilizations may
appear within a thin spacelike slab running along the interior reheating
surface. This reheating surface appears to interior observers as an
infinite, spacelike hypersurface~\cite{Coleman:1980aw}. For this
reason, such bubbles are called {}``pocket universes,'' while the
spacetime is called a {}``multiverse.'' (Generally, the term {}``pocket
universe'' refers to a noncompact, connected component of the reheating
surface~\cite{Guth:1999rh}.)

In scenarios of this type, each bubble is causally disconnected from
most other bubbles.%
\footnote{Collisions between bubbles are rare~\cite{Guth:1982pn}; however,
effects of bubble collisions are observable in principle~\cite{Garriga:2006hw}.%
} Hence, bubble nucleation events may generate infinitely many statistically
inequivalent, causally disconnected patches of the reheating surface,
every patch giving rise to a possibly infinite number of galaxies
and observers. This feature significantly complicates the task of
extracting physical predictions from these models. This class of models
is referred to as {}``eternal inflation of tunneling type.''

In the following subsections, I discuss the motivation for studying
eternal inflation as well as physical justifications for adopting
the effective stochastic picture. Different techniques developed for
describing eternal inflation are reviewed in Sec.~\ref{sec:Stochastic approach}.
Section~\ref{sec:Predictions-and-cutoffs} contains an overview of
methods for extracting predictions and a discussion of the accompanying
{}``measure problem.''

\subsection{Some motivation\label{sub:Physical-relevance}}

The hypothesis of cosmological inflation was invoked to explain several
outstanding puzzles in observational data~\cite{Guth:1980zm}. However,
some observed quantities (such as the cosmological constant $\Lambda$
or elementary particle masses) may be expectation values of slowly-varying
effective fields $\chi_{a}$. Within the phenomenological approach,
we are compelled to consider also the fluctuations of the fields $\chi_{a}$
during inflation, on the same footing as the fluctuations of the inflaton
$\phi$. Hence, in a generic scenario of eternal inflation, all the
fields $\chi_{a}$ arrive at the reheating surface $\phi=\phi_{*}$
with values that can be determined only statistically. Observers appearing
at different points in space may thus measure different values of
the cosmological constant, elementary particle masses, spectra of
primordial density fluctuations, and other cosmological parameters.

It is important to note that inhomogeneities in observable quantities
are created on scales far exceeding the Hubble horizon scale. Such
inhomogeneities are not directly accessible to astrophysical experiments.
Nevertheless, the study of the global structure of eternally inflating
spacetime is not merely of academic interest. Fundamental questions
regarding the cosmological singularities, the beginning of the Universe
and of its ultimate fate, as well as the issue of the cosmological
initial conditions all depend on knowledge of the global structure
of the spacetime as predicted by the theory, whether or not this global
structure is directly observable (see e.g.~\cite{Garriga:1999hf,Garriga:2001ch}).
In other words, the fact that some theories predict eternal inflation
influences our assessment of the viability of these theories. In particular,
the problem of initial conditions for inflation~\cite{Goldwirth:1991rj}
is significantly alleviated when eternal inflation is present. For
instance, it was noted early on that the presence of eternal self-reproduction
in the {}``chaotic'' inflationary scenario~\cite{Linde:1983gd}
essentially removes the need for the fine-tuning of the initial conditions~\cite{Linde:1986fc,Linde:1986fd}.
More recently, constraints on initial conditions were studied in the
context of self-reproduction in models of quintessence~\cite{Martin:2004ba}
and $k$-inflation~\cite{Helmer:2006tz}.

Since the values of the observable parameters $\chi_{a}$ are random,
it is natural to ask for the probability distribution of $\chi_{a}$
that would be measured by a randomly chosen observer. Understandably,
this question has been the main theme of much of the work on eternal
inflation. Obtaining an answer to this question promises to establish
a more direct contact between scenarios of eternal inflation and experiment.
For instance, if the probability distribution for the cosmological
constant $\Lambda$ were peaked near the experimentally observed,
puzzlingly small value (see e.g.~\cite{Carroll:2000fy} for a review
of the cosmological constant problem), the smallness of $\Lambda$
would be explained as due to observer selection effects rather than
to fundamental physics. Considerations of this sort necessarily involve
some anthropic reasoning; however, the relevant assumptions are minimal.
The basic goal of theoretical cosmology is to select physical theories
of the early universe that are most compatible with astrophysical
observations, including the observation of our existence. It appears
reasonable to assume that the civilization of Planet Earth evolved
near a randomly chosen star compatible with the development of life,
within a randomly chosen galaxy where such stars exist. Many models
of inflation generically include eternal inflation and hence predict
the formation of infinitely many galaxies where civilizations like
ours may develop. It is then also reasonable to assume that our civilization
is typical among all the civilizations that evolved in galaxies formed
at any time in the universe. This assumption is called the {}``principle
of mediocrity''~\cite{Vilenkin:1995yd}.

To use the {}``principle of mediocrity'' for extracting statistical
predictions from a model of eternal inflation, one proceeds as follows~\cite{Vilenkin:1995yd,Vilenkin:1998kr}.
In the example with the fields $\chi_{a}$ described above, the question
is to determine the probability distribution for the values of $\chi_{a}$
that a random observer will measure. Presumably, the values of the
fields $\chi_{a}$ do not directly influence the emergence of intelligent
life on planets, although they may affect the efficiency of structure
formation or nucleosynthesis. Therefore, we may assume a fixed, $\chi_{a}$-dependent
mean number of civilizations $\nu_{\text{civ}}(\chi_{a})$ per galaxy
and proceed to ask for the probability distribution $P_{G}(\chi_{a})$
of $\chi_{a}$ near a randomly chosen galaxy. The observed probability
distribution of $\chi_{a}$ will then be\begin{equation}
P(\chi_{a})=P_{G}(\chi_{a})\nu_{\text{civ}}(\chi_{a}).\end{equation}
 One may use the standard {}``hot big bang'' cosmology to determine
the average number $\nu_{G}(\chi_{a})$ of suitable galaxies per unit
volume in a region where reheating occurred with given values of $\chi_{a}$;
in any case, this task does not appear to pose difficulties of principle.
Then the computation of $P_{G}(\chi_{a})$ is reduced to determining
the volume-weighted probability distribution $\mathcal{V}(\chi_{a})$
for the fields $\chi_{a}$ within a randomly chosen 3-volume along
the reheating surface. The probability distribution of $\chi_{a}$
will be expressed as\begin{equation}
P(\chi_{a})=\mathcal{V}(\chi_{a})\nu_{G}(\chi_{a})\nu_{\text{civ}}(\chi_{a}).\label{eq:probability}\end{equation}
However, defining $\mathcal{V}(\chi_{a})$ turns out to be far from
straightforward since the reheating surface in eternal inflation is
an infinite 3-surface with a complicated geometry and topology. The
lack of a natural, unambiguous, unbiased measure on the infinite reheating
surface is known as the {}``measure problem'' in eternal inflation.
Existing approaches and measure prescriptions are discussed in Sec.~\ref{sec:Predictions-and-cutoffs},
where two main alternatives (the {}``volume-based'' and {}``worldline-based''
measures) are presented. In Sections~\ref{sub:Predictions-of-observable}
and \ref{sub:Predictions-in-discrete} I give arguments in favor of
using the volume-based measure for computing the probability distribution
of values $\chi_{a}$ measured by a random observer. The volume-based
measure has been applied to obtain statistical predictions for the
gravitational constant in Brans-Dicke theories~\cite{Garcia-Bellido:1993wn,Garcia-Bellido:1994vz},
cosmological constant (dark energy)~\cite{Garriga:1998px,Garriga:1999bf,Garriga:1999hu,Garriga:2002tq,Garriga:2003hj,Garriga:2005ee},
particle physics parameters~\cite{Tegmark:2003ug,Tegmark:2005dy,Hall:2006ff},
and the amplitude of primordial density perturbations~\cite{Susperregi:1996hs,Garriga:1999hu,Feldstein:2005bm,Garriga:2005ee}.

The issue of statistical predictions has recently come to the fore
in conjunction with the discovery of the string theory landscape.
According to various estimates, one expects to have between $10^{500}$
and $10^{1500}$ possible vacuum states of string theory~\cite{Lerche:1986cx,Bousso:2000xa,Susskind:2003kw,Kachru:2003aw,Denef:2004ze}.
The string vacua differ in the geometry of spacetime compactification
and have different values of the effective cosmological constant (or
{}``dark energy'' density). Transitions between vacua may happen
via the well-known Coleman-deLuccia tunneling mechanism~\cite{Coleman:1980aw}.
Once the dark energy dominates in a given region, the spacetime becomes
locally de Sitter. Then the tunneling process will create infinitely
many disconnected {}``daughter'' bubbles of other vacua. Observers
like us may appear within any of the habitable bubbles. Since the
fundamental theory does not specify a single {}``preferred'' vacuum,
it remains to try determining the probability distribution of vacua
as found by a randomly chosen observer. The {}``volume-based'' and
{}``worldline-based'' measures can be extended to scenarios with
multiple bubbles, as discussed in more detail in Sec.~\ref{sub:Predictions-in-discrete}.
Some recent results obtained using these measures are reported in
Refs.~\cite{Schwartz-Perlov:2006hi,Aguirre:2006ak,Linde:2006nw}.

\subsection{Physical justifications of the semiclassical picture\label{sub:Physical-justifications}}

The standard framework of inflationary cosmology asserts that vacuum
quantum fluctuations with super-horizon wavelengths become classical
inhomogeneities of the field $\phi$. The calculations of cosmological
density perturbations generated during inflation~\cite{Mukhanov:1981xt,Mukhanov:1982nu,Hawking:1982cz,Starobinsky:1982ee,Guth:1982ec,Bardeen:1983qw,Mukhanov:1990me}
also assume that a {}``classicalization'' of quantum fluctuations
takes place via the same mechanism. In the calculations, the statistical
average $\left\langle \delta\phi^{2}\right\rangle $ of classical
fluctuations on super-Hubble scales is simply set equal to the quantum
expectation value $\left\langle 0\right|\hat{\phi}^{2}\left|0\right\rangle $
in a suitable vacuum state.  While this approach is widely accepted
in the cosmology literature, a growing body of research is devoted
to the analysis of the quantum-to-classical transition during inflation
(see e.g.~\cite{Brandenberger:1984cz} for an early review). Since
a detailed analysis would be beyond the scope of the present text,
I merely outline the main ideas and arguments relevant to this issue.

A standard phenomenological explanation of the {}``classicalization''
of the perturbations is as follows. For simplicity, let us restrict
our attention to a slow-roll inflationary scenario with one scalar
field $\phi$. In the slow roll regime, one can approximately regard
$\phi$ as a massless scalar field in de Sitter background spacetime~\cite{Goncharov:1987ir}.
Due to the exponentially fast expansion of de Sitter spacetime, super-horizon
Fourier modes of the field $\phi$ are in squeezed quantum states
with exponentially large ($\sim e^{Ht}$) squeezing parameters~\cite{Polarski:1995jg,Kiefer:1998jk,Kiefer:1998pb,Kiefer:1998qe,Kiefer:1999sj,Kiefer:2006je}.
Such highly squeezed states have a macroscopically large uncertainty
in the field value $\phi$ and thus quickly decohere due to interactions
with gravity and with other fields. The resulting mixed state is effectively
equivalent to a statistical ensemble with a Gaussian distributed value
of $\phi$. Therefore one may compute the statistical average $\left\langle \delta\phi^{2}\right\rangle $
as the quantum expectation value $\left\langle 0\right|\hat{\phi}^{2}\left|0\right\rangle $
and interpret the fluctuation $\delta\phi$ as a classical {}``noise.''
A heuristic description of the {}``classicalization''~\cite{Goncharov:1987ir}
is that the quantum commutators of the creation and annihilation operators
of the field modes, $[\hat{a},\hat{a}^{\dagger}]=1$, are much smaller
than the expectation values $\left\langle a^{\dagger}a\right\rangle \gg1$
and are thus negligible.

A related issue is the backreaction of fluctuations of the scalar
field $\phi$ on the metric.%
\footnote{The backreaction effects of the long-wavelength fluctuations of a
scalar field during inflation have been investigated extensively (see
e.g.~\cite{Abramo:1997hu,Afshordi:2000nr,Abramo:2001db,Abramo:2001dc,Abramo:2001dd,Geshnizjani:2003cn,Geshnizjani:2004tf,Kahya:2006hc}).%
} According to the standard theory (see e.g.~\cite{Kodama:1985bj,Mukhanov:1990me}
for reviews), the perturbations of the metric arising due to fluctuations
of $\phi$ are described by an auxiliary scalar field (sometimes called
the {}``Sasaki-Mukhanov variable'') in a fixed de Sitter background.
Thus, the {}``classicalization'' effect should apply equally to
the fluctuations of $\phi$ and to the induced metric perturbations.
At the same time, these metric perturbations can be viewed, in an
appropriate coordinate system, as fluctuations of the local expansion
rate $H(\phi)$ due to local fluctuations of $\phi$~\cite{Starobinsky:1982ee,Goncharov:1987ir,Frolov:2002va}.
Thus one arrives at the picture of a {}``locally de Sitter'' spacetime
with the metric~(\ref{eq:locally de Sitter}), where the Hubble rate
$\dot{a}/a=H(\phi)$ fluctuates on super-horizon length scales and
locally follows the value of $\phi$ via the classical Einstein equation~(\ref{eq:Einstein1}).

The picture as outlined is phenomenological and does not provide a
description of the quantum-to-classical transition in the metric perturbations
at the level of field theory. For instance, a fluctuation of $\phi$
leading to a local increase of $H(\phi)$ necessarily violates the
null energy condition~\cite{Borde:1997pp,Winitzki:2001fc,Vachaspati:2003de}.
The cosmological implications of such {}``semiclassical'' fluctuations
(see e.g.~the scenario of {}``island cosmology''~\cite{Dutta:2005gt,Piao:2005na,Dutta:2005if})
cannot be understood in detail within the framework of the phenomenological
picture.

A more fundamental approach to describing the quantum-to-classical
transition of perturbations was developed using non-equilibrium quantum
field theory and the influence functional formalism~\cite{Morikawa:1987ci,Morikawa:1989xz,Hu:1994ep,Calzetta:1995ys}.
In this approach, decoherence of a pure quantum state of $\phi$ into
a mixed state is entirely due to the self-interaction of the field
$\phi$. In particular, it is predicted that no decoherence would
occur for a free field with $V(\phi)=\frac{1}{2}m\phi^{2}$. This
result is at variance with the accepted paradigm of {}``classicalization''
as outlined above. If the source of the {}``noise'' is the coupling
between different perturbation modes of $\phi$, the typical amplitude
of the {}``noise'' will be second-order in the perturbation. This
is several orders of magnitude smaller than the amplitude of {}``noise''
found in the standard approach. Accordingly, it is claimed~\cite{Matacz:1996gk,Matacz:1996fv}
that the magnitude of cosmological perturbations generated by inflation
is several orders of magnitude smaller than the results currently
accepted as standard, and that the shape of the perturbation spectrum
depends on the details of the process of {}``classicalization''~\cite{Kubotani:1997xa}.
Thus, the results obtained via the influence functional techniques
do not appear to reproduce the phenomenological picture of {}``classicalization''
as outlined above. This mismatch emphasizes the need for a deeper
understanding of the nature of the quantum-to-classical transition
for cosmological perturbations.

Finally, let us mention a different line of work which supports the
{}``classicalization'' picture. In Refs.~\cite{Vilenkin:1983xp,Woodard:2005cw,Tsamis:2005hd,Woodard:2006pw,Miao:2006pn,Prokopec:2006ue},
calculations of (renormalized) expectation values such as $\langle\hat{\phi}^{2}\rangle$,
$\langle\hat{\phi}^{4}\rangle$, etc., were performed for field operators
$\hat{\phi}$ in a fixed de Sitter background. The results were compared
with the statistical averages\begin{equation}
\int P(\phi,t)\phi^{2}d\phi,\quad\int P(\phi,t)\phi^{4}d\phi,\quad\text{etc.},\label{eq:averages phi}\end{equation}
where the distribution $P(\phi,t)$ describes the {}``random walk''
of the field $\phi$ in the Fokker-Plank approach (see Sec.~\ref{sub:Eternal-inflation-from}).
It was shown that the leading late-time asymptotics of the quantum
expectation values coincide with the corresponding statistical averages~(\ref{eq:averages phi}).
These results appear to validate the {}``random walk'' approach,
albeit in a limited context (in the absence of backreaction).

\section{Stochastic approach to inflation \label{sec:Stochastic approach}}

The \emph{stochastic approach to inflation} is a semiclassical, statistical
description of the spacetime resulting from quantum fluctuations of
the inflaton field(s) and their backreaction on the metric~\cite{Vilenkin:1983xq,Starobinsky:1986fx,Rey:1986zk,Nakao:1988yi,Kandrup:1988sc,Nambu:1988je,Nambu:1989uf,Mijic:1990qx,Salopek:1990re,Linde:1993nz,Linde:1993xx}.
In this description, the spacetime remains everywhere classical 
but its geometry is determined by a stochastic process. In the next
subsections I review the main tools used in the stochastic approach
for calculations in the context of random-walk type, slow-roll inflation.
Models involving tunneling-type eternal inflation are considered in
Sec.~\ref{sub:Bubble-nucleation-models}.

\subsection{Random walk-type eternal inflation\label{sub:Eternal-inflation-from}}

The important features of random walk-type eternal inflation can be
understood by considering a simple slow-roll inflationary model with
a single scalar field $\phi$ and a potential $V(\phi)$. The slow-roll
evolution equation is\begin{equation}
\dot{\phi}=-\frac{1}{3H}\frac{dV}{d\phi}=-\frac{1}{4\pi}\frac{dH}{d\phi}\equiv v(\phi),\label{eq:v def}\end{equation}
where $H(\phi)$ is defined by Eq.~(\ref{eq:Einstein1}) and $v(\phi)$
is a model-dependent function describing the {}``velocity'' $\dot{\phi}$
of the deterministic evolution of the field $\phi$. The slow-roll
trajectory $\phi_{\text{sr}}(t)$, which is a solution of Eq.~(\ref{eq:v def}),
is an attractor~\cite{Salopek:1990jq,Liddle:1994dx} for trajectories
starting with a wide range of initial conditions.%
\footnote{See Ref.~\cite{Helmer:2006tz} for a precise definition of an attractor
trajectory in the context of inflation. %
}

As discussed in Sec.~\ref{sub:Physical-justifications}, the super-horizon
modes of the field $\phi$ are assumed to undergo a rapid quantum-to-classical
transition. Therefore one regards the spatial average of $\phi$ on
scales of several $H^{-1}$ as a \emph{classical} field variable.
The spatial averaging can be described with help of a suitable window
function,\begin{equation}
\left\langle \phi(\mathbf{x})\right\rangle \equiv\int W(\mathbf{x}-\mathbf{y})\phi(\mathbf{y})d^{3}\mathbf{y}.\end{equation}
It is implied that the window function $W(\mathbf{x})$ decays quickly
on physical distances $a\left|\mathbf{x}\right|$ of order several
$H^{-1}$. From now on, let us denote the volume-averaged field simply
by $\phi$ (no other field $\phi$ will be used). 

As discussed above, the influence of quantum fluctuations leads to
random {}``jumps'' superimposed on top of the deterministic evolution
of the volume-averaged field $\phi(t,\mathbf{x})$. This may be described
by a Langevin equation of the form~\cite{Starobinsky:1986fx}\begin{equation}
\dot{\phi}(t,\mathbf{x})=v(\phi)+N(t,\mathbf{x}),\label{eq:Langevin}\end{equation}
where $N(t,\mathbf{x})$ stands for {}``noise'' and is assumed to
be a Gaussian random function with known correlator~\cite{Starobinsky:1986fx,Bellini:1996uh,Casini:1998wr,Winitzki:1999ve}
\begin{equation}
\left\langle N(t,\mathbf{x})N(\tilde{t},\tilde{\mathbf{x}})\right\rangle =C(t,\tilde{t},\left|\mathbf{x}-\tilde{\mathbf{x}}\right|;\phi).\end{equation}
An explicit form of the correlator $C$ depends on the specific window
function $W$ used for averaging the field $\phi$ on Hubble scales~\cite{Winitzki:1999ve}.
However, the window function $W$ is merely a phenomenological device
used in lieu of a complete \emph{ab initio} derivation of the stochastic
inflation picture. One expects, therefore, that results of calculations
should be robust with respect to the choice of $W$. In other words,
any uncertainty due to the choice of the window function must be regarded
as an imprecision inherent in the method. For instance, a robust result
in this sense is an exponentially fast decay of correlations on time
scales $\Delta t\gtrsim H^{-1}$,\begin{equation}
C(t,\tilde{t},\left|\mathbf{x}-\tilde{\mathbf{x}}\right|;\phi)\propto\exp\left(-2H(\phi)\left|t-\tilde{t}\right|\right),\label{eq:C decay}\end{equation}
which holds for a wide class of window functions~\cite{Winitzki:1999ve}.

For the purposes of the present consideration, we only need to track
the evolution of $\phi(t,\mathbf{x})$ along a single comoving worldline
$\mathbf{x}=\text{const}$. Thus, we will not need an explicit form
of $C(t,\tilde{t},\left|\mathbf{x}-\tilde{\mathbf{x}}\right|;\phi)$
but merely the value at coincident points $t=\tilde{t}$, $\mathbf{x}=\tilde{\mathbf{x}}$,
which is computed in the slow-roll inflationary scenario as~\cite{Starobinsky:1986fx}\begin{equation}
C(t,t,0;\phi)=\frac{H^{2}(\phi)}{4\pi^{2}}.\end{equation}
(This represents the fluctuation~(\ref{eq:delta-phi}) accumulated
during one Hubble time, $\Delta t=H^{-1}$.) Due to the property~(\ref{eq:C decay}),
one may neglect correlations on time scales $\Delta t\gtrsim H^{-1}$
in the {}``noise'' field.%
\footnote{Taking these correlations into account leads to a picture of {}``color
noise''~\cite{Matarrese:2003ye,Liguori:2004fa}. In what follows,
we only consider the simpler picture of {}``white noise'' as an
approximation adequate for the issues at hand.%
} Thus, the evolution of $\phi$ on time scales $\Delta t\gtrsim H^{-1}$
can be described by a finite-difference form of the Langevin equation~(\ref{eq:Langevin}),\begin{equation}
\phi(t+\Delta t)-\phi(t)=v(\phi)\Delta t+\sqrt{2D(\phi)\Delta t}\,\xi(t),\label{eq:Langevin FD}\end{equation}
where\begin{equation}
D(\phi)\equiv\frac{H^{3}(\phi)}{8\pi^{2}}\label{eq:D def}\end{equation}
 and $\xi$ is a normalized random variable representing {}``white
noise,''\begin{align}
\left\langle \xi\right\rangle  & =0,\quad\left\langle \xi^{2}\right\rangle =1,\\
\left\langle \xi(t)\xi(t+\Delta t)\right\rangle  & =0\quad\text{for}\quad\Delta t\gtrsim H^{-1}.\end{align}

Equation~(\ref{eq:Langevin FD}) is interpreted as describing a Brownian
motion $\phi(t)$ with the systematic {}``drift'' $v(\phi)$ and
the {}``diffusion coefficient'' $D(\phi)$. In a typical slow-roll
inflationary scenario, there will be a range of $\phi$ where the
noise dominates over the deterministic drift, \begin{equation}
v(\phi)\Delta t\ll\sqrt{2D(\phi)\Delta t},\quad\Delta t\equiv H^{-1}.\label{eq:diffusion dominated}\end{equation}
 Such a range of $\phi$ is called the {}``diffusion-dominated regime.''
For $\phi$ near the end of inflation, the amplitude of the noise
is very small, and so the opposite inequality holds. This is the {}``deterministic
regime'' where the random jumps can be neglected and the field $\phi$
follows the slow-roll trajectory.

\subsection{Fokker-Planck equations\label{sub:Fokker-Planck-equations}}

A useful description of the statistical properties of $\phi(t)$ is
furnished by the probability density $P(\phi,t)d\phi$ of having a
value $\phi$ at time $t$. As in the case of the Langevin equation,
the values $\phi(t)$ are measured along a single, randomly chosen
comoving worldline $\mathbf{x}=\text{const}$. The probability distribution
$P(\phi,t)$ satisfies the Fokker-Planck (FP) equation whose standard
derivation we omit~\cite{vanKampen:1981:SPiPC,Risken:1989:TFPE},\begin{equation}
\partial_{t}P=\partial_{\phi}\left[-v(\phi)P+\partial_{\phi}\left(D(\phi)P\right)\right].\label{eq:FP c}\end{equation}
 The coefficients $v(\phi)$ and $D(\phi)$ are in general model-dependent
and need to be calculated in each particular scenario. These calculations
require only the knowledge of the slow-roll trajectory and the mode
functions of the quantized scalar perturbations. For ordinary slow-roll
inflation with an effective potential $V(\phi)$, the results are
well-known expressions~(\ref{eq:v def}) and~(\ref{eq:D def}).
The corresponding expressions for models of $k$-inflation were derived
in Ref.~\cite{Helmer:2006tz} using the relevant quantum theory of
perturbations~\cite{Garriga:1999vw}. 

It is well known that there exists a {}``factor ordering'' ambiguity
in translating the Langevin equation into the FP equation if the amplitude
of the {}``noise'' depends on the position. Specifically, the factor
$D(\phi)$ in Eq.~(\ref{eq:Langevin FD}) may be replaced by $D(\phi+\theta\Delta t)$,
where $0<\theta<1$ is an arbitrary constant. With $\theta\neq0$,
the term $\partial_{\phi\phi}\left(DP\right)$ in Eq.~(\ref{eq:FP c})
will be replaced by a different ordering of the factors, \begin{equation}
\partial_{\phi\phi}\left(DP\right)\rightarrow\partial_{\phi}\left[D^{\theta}\partial_{\phi}\left(D^{1-\theta}P\right)\right].\end{equation}
 Popular choices $\theta=0$ and $\theta=\frac{1}{2}$ are called
the Ito and the Stratonovich factor ordering respectively. Motivated
by the considerations of Ref.~\cite{Vilenkin:1999kd}, we choose
$\theta=0$ as shown in Eqs.~(\ref{eq:Langevin FD}) and (\ref{eq:FP c}).
Given the phenomenological nature of the Langevin equation~(\ref{eq:Langevin FD}),
one expects that any ambiguity due to the choice of $\theta$ represents
an imprecision inherent in the stochastic approach. This imprecision
is typically of order $H^{2}\ll1$~\cite{Winitzki:1995pg}.

The quantity $P(\phi,t)$ may be also interpreted as the fraction
of the \emph{comoving} volume (i.e.~coordinate volume $d^{3}\mathbf{x}$)
occupied by the field value $\phi$ at time $t$. Another important
characteristic is the volume-weighted distribution $P_{V}(\phi,t)d\phi$,
which is defined as the \emph{proper} 3-volume (as opposed to the
comoving volume) of regions having the value $\phi$ at time $t$.
(To avoid considering infinite volumes, one may restrict one's attention
to a finite comoving domain in the universe and normalize $P_{V}(\phi,t)$
to unit volume at some initial time $t=t_{0}$.) The volume distribution
satisfies a modified FP equation~\cite{Goncharov:1987ir,Nambu:1988je,Mijic:1990qx},\begin{equation}
\partial_{t}P_{V}=\partial_{\phi}\left[-v(\phi)P_{V}+\partial_{\phi}\left(D(\phi)P_{V}\right)\right]+3H(\phi)P_{V},\label{eq:FP V}\end{equation}
 which differs from Eq.~(\ref{eq:FP c}) by the term $3HP_{V}$ that
describes the exponential growth of 3-volume in inflating regions.%
\footnote{A more formal derivation of Eq.~(\ref{eq:FP V}) as well as details
of the interpretation of the distributions $P$ and $P_{V}$ in terms
of ensembles of worldlines can be found in Ref.~\cite{Winitzki:2005ya}.%
} 

Presently we consider scenarios with a single scalar field; however,
the formalism of FP equations can be straightforwardly extended to
multi-field models (see e.g.~Ref.~\cite{Vilenkin:1999kd}). For
instance, the FP equation for a two-field model is\begin{equation}
\partial_{t}P=\partial_{\phi\phi}\left(DP\right)+\partial_{\chi\chi}\left(DP\right)-\partial_{\phi}\left(v_{\phi}P\right)-\partial_{\chi}\left(v_{\chi}P\right),\label{eq:FP 2 field}\end{equation}
where $D(\phi,\chi)$, $v_{\phi}(\phi,\chi)$, and $v_{\chi}(\phi,\chi)$
are appropriate coefficients.

\subsection{Methods of solution\label{sub:Methods-of-solution}}

In principle, one can solve the FP equations forward in time by a
numerical method, starting from a given initial distribution at $t=t_{0}$.
To specify the solution uniquely, the FP equations must be supplemented
by boundary conditions at both ends of the inflating range of $\phi$~\cite{Linde:1993nz,Linde:1993xx}.
At the reheating boundary ($\phi=\phi_{*}$), one imposes the {}``exit''
boundary conditions,\begin{equation}
\partial_{\phi}\left[D(\phi)P\right]_{\phi=\phi_{*}}=0,\quad\partial_{\phi}\left[D(\phi)P_{V}\right]_{\phi=\phi_{*}}=0.\label{eq:exit BC}\end{equation}
These boundary conditions express the fact that random jumps are very
small at the end of inflation and cannot move the value of $\phi$
away from $\phi=\phi_{*}$. If the potential $V(\phi)$ reaches Planck
energy scales at some $\phi=\phi_{\max}$ (this happens generally
in {}``chaotic'' type inflationary scenarios with unbounded potentials),
the semiclassical picture of spacetime breaks down for regions with
$\phi\sim\phi_{\max}$. Hence, a boundary condition must be imposed
also at $\phi=\phi_{\max}$. For instance, one can use the absorbing
boundary condition,\begin{equation}
P(\phi_{\max})=0,\end{equation}
which means that Planck-energy regions with $\phi=\phi_{\max}$ disappear
from consideration~\cite{Linde:1993nz,Linde:1993xx}. 

Once the boundary conditions are specified, one may write the general
solution of the FP equation~(\ref{eq:FP c}) as\begin{equation}
P(\phi,t)=\sum_{\lambda}C_{\lambda}P^{(\lambda)}(\phi)\, e^{\lambda t},\end{equation}
where the sum is performed over all the eigenvalues $\lambda$ of
the differential operator\begin{equation}
\hat{L}P\equiv\partial_{\phi}\left[-v(\phi)P+\partial_{\phi}\left(D(\phi)P\right)\right],\end{equation}
and the corresponding eigenfunctions $P^{(\lambda)}$ are defined
by\begin{equation}
\hat{L}P^{(\lambda)}(\phi)=\lambda P^{(\lambda)}(\phi).\end{equation}
 The constants $C_{\lambda}$ can be expressed through the initial
distribution $P(\phi,t_{0})$.

By an appropriate change of variables $\phi\rightarrow z$, $P(\phi)\rightarrow F(z)$,
the operator $\hat{L}$ may be brought into a manifestly self-adjoint
form~\cite{Nakao:1988yi,Nambu:1988je,Matarrese:1988hc,Nambu:1989uf,Mijic:1990qx,Winitzki:1995pg},\begin{equation}
\hat{L}\,\rightarrow\,\frac{d^{2}}{dz^{2}}+U(z).\end{equation}
 Then one can show that all the eigenvalues $\lambda$ of $\hat{L}$
are nonpositive; in particular, the (algebraically) largest eigenvalue
$\lambda_{\max}\equiv-\gamma<0$ is nondegenerate and the corresponding
eigenfunction $P^{(\lambda_{\max})}(\phi)$ is everywhere positive~\cite{Winitzki:1995pg,Helmer:2006tz}.
Hence, this eigenfunction describes the late-time asymptotic of the
distribution $P(\phi,t)$,\begin{equation}
P(\phi,t)\propto P^{(\lambda_{\max})}(\phi)\, e^{-\gamma t}.\label{eq:late-time c}\end{equation}
 The distribution $P^{(\lambda_{\max})}(\phi)$ is the {}``stationary''
distribution of $\phi$ per comoving volume at late times. The exponential
decay of the distribution $P(\phi,t)$ means that at late times most
of the comoving volume (except for an exponentially small fraction)
has finished inflation and entered reheating.

Similarly, one can represent the general solution of Eq.~(\ref{eq:FP V})
by\begin{equation}
P_{V}(\phi,t)=\sum_{\tilde{\lambda}}C_{\tilde{\lambda}}P_{V}^{(\tilde{\lambda})}(\phi)e^{\tilde{\lambda}t},\end{equation}
where\begin{equation}
[\hat{L}+3H(\phi)]P^{(\tilde{\lambda})}(\phi)=\tilde{\lambda}P^{(\tilde{\lambda})}(\phi).\end{equation}
By the same method as for the operator $\hat{L}$, it is possible
to show that the spectrum of eigenvalues $\tilde{\lambda}$ of the
operator $\hat{L}+3H(\phi)$ is bounded from above and that the largest
eigenvalue $\tilde{\lambda}_{\max}\equiv\tilde{\gamma}$ admits a
nondegenerate, everywhere positive eigenfunction $P^{(\tilde{\gamma})}(\phi)$.
However, the largest eigenvalue $\tilde{\gamma}$ may be either positive
or negative. If $\tilde{\gamma}>0$, the late-time behavior of $P_{V}(\phi,t)$
is\begin{equation}
P_{V}(\phi,t)\propto P^{(\tilde{\gamma})}(\phi)e^{\tilde{\gamma}t},\label{eq:late-time V}\end{equation}
which means that the total proper volume of all the inflating regions
grows with time. This is the behavior expected in eternal inflation:
the number of independently inflating domains increases without limit.
Thus, the condition $\tilde{\gamma}>0$ is the criterion for the presence
of eternal self-reproduction of inflating domains. The corresponding
distribution $P^{(\tilde{\gamma})}(\phi)$ is called the {}``stationary''
distribution~\cite{Linde:1993nz,Linde:1993xx,Garcia-Bellido:1994ci,Garcia-Bellido:1995kc}. 

If $\tilde{\gamma}\leq0$, eternal inflation does not occur and the
entire space almost surely (i.e.~with probability 1) enters the reheating
epoch at a finite time.

If the potential $V(\phi)$ is of {}``new'' inflationary type~\cite{Linde:1981mu,Linde:1982zj,Linde:1982uu,Linde:1982tg,Albrecht:1982wi}
and has a global maximum at say $\phi=\phi_{0}$, the eigenvalues
$\gamma$ and $\tilde{\gamma}$ can be estimated (under the usual
slow-roll assumptions on $V$) as~\cite{Winitzki:1995pg}\begin{equation}
\gamma\approx\frac{V^{\prime\prime}(\phi_{0})}{8\pi V(\phi_{0})}H(\phi_{0})<0,\quad\tilde{\gamma}\approx3H(\phi_{0})>0.\end{equation}
Therefore, eternal inflation is generic in the {}``new'' inflationary
scenario.

Let us comment on the possibility of obtaining solutions $P(\phi,t)$
in practice. With the potential $V(\phi)=\lambda\phi^{4}$, the full
time-dependent FP equation~(\ref{eq:FP c}) can be solved analytically
via a nonlinear change of variable $\phi\rightarrow\phi^{-2}$~\cite{Hodges:1989zz,Matarrese:1988hc,Yi:1991ub}.
This exact solution, as well as an approximate solution $P(\phi,t)$
for a general potential, can be also obtained using the saddle-point
evaluation of a path-integral expression for $P(\phi,t)$~\cite{Shtanov:1994nt}.
In some cases the eigenvalue equation $\hat{L}P^{(\lambda)}=\lambda P^{(\lambda)}$
may be reduced to an exactly solvable Schr\"odinger equation. These
cases include potentials of the form $V(\phi)=\lambda e^{\mu\phi}$,
$V(\phi)=\lambda\phi^{-2}$, $V(\phi)=\lambda\cosh^{-2}(\mu\phi)$;
see e.g.~\cite{Winitzki:1995pg} for other examples. 

A general approximate method for determining $P(\phi,t)$ for arbitrary
potentials~\cite{Martin:2005ir,Gratton:2005bi,Martin:2005hb} consists
of a perturbative expansion,\begin{equation}
\phi(t)=\phi_{0}(t)+\delta\phi_{1}(t)+\delta\phi_{2}(t)+...,\end{equation}
 applied directly to the Langevin equation. The result is (at the
lowest order) a Gaussian approximation with a time-dependent mean
and variance~\cite{Martin:2005ir},\begin{align}
P(\phi,t) & \approx\frac{1}{\sqrt{2\pi\sigma^{2}(t)}}\exp\left[-\frac{\left(\phi-\phi_{0}(t)\right)^{2}}{2\sigma^{2}(t)}\right],\label{eq:P Gaussian}\\
\sigma^{2}(t) & \equiv\frac{H^{\prime2}(\phi_{\text{sr}})}{\pi}\int_{\phi_{\text{sr}}}^{\phi_{\text{in}}}\frac{H^{3}}{H^{\prime3}}d\phi,\\
\phi_{0}(t) & \equiv\phi_{\text{sr}}(t)+\frac{H^{\prime\prime}}{2H^{\prime}}\sigma^{2}(t)+\frac{H^{\prime}}{4\pi}\left[\frac{H_{\text{in}}^{3}}{H_{\text{in}}^{\prime2}}-\frac{H^{3}}{H^{\prime2}}\right],\end{align}
where $\phi_{\text{sr}}(t)$ is the slow-roll trajectory and $\phi_{\text{in}}$
is the initial value of $\phi$. While methods based on the Langevin
equation do not take into account boundary conditions or volume weighting
effects, the formula~(\ref{eq:P Gaussian}) provides an adequate
approximation to the distribution $P(\phi,t)$ in a useful range of
$\phi$ and $t$~\cite{Martin:2005hb}.

\subsection{Gauge dependence issues\label{sub:Gauge-dependence-issues}}

An important feature of the FP equations is their dependence on the
choice of the time variable. One can consider a replacement of the
form\begin{equation}
t\rightarrow\tau,\quad d\tau\equiv T(\phi)dt,\label{eq:t tau}\end{equation}
understood in the sense of integrating along comoving worldlines $\mathbf{x}=\text{const}$,
where $T(\phi)>0$ is an arbitrary function of the field. For instance,
a possible choice is $T(\phi)\equiv H(\phi)$, which makes the new
time variable dimensionless, \begin{equation}
\tau=\int Hdt=\ln a.\end{equation}
This time variable is called {}``scale factor time'' or {}``$e$-folding
time'' since it measures the number of $e$-foldings along a comoving
worldline. 

The distributions $P(\phi,\tau)$ and $P_{V}(\phi,\tau)$ are defined
as before, except for considering the 3-volumes along hypersurfaces
of equal $\tau$. These distributions satisfy FP equations similar
to Eqs.~(\ref{eq:FP c})--(\ref{eq:FP V}). With the replacement~(\ref{eq:t tau}),
the coefficients of the new FP equations are modified as follows~\cite{Winitzki:1995pg},\begin{equation}
D(\phi)\rightarrow\frac{D(\phi)}{T(\phi)},\quad v(\phi)\rightarrow\frac{v(\phi)}{T(\phi)},\end{equation}
while the {}``growth'' term $3HP_{V}$ in Eq.~(\ref{eq:FP V})
is replaced by $3HT^{-1}P_{V}$. The change in the coefficients may
significantly alter the qualitative behavior of the solutions of the
FP equations. For instance, stationary distributions defined through
the proper time $t$ and the $e$-folding time $\tau=\ln a$ were
found to have radically different behavior~\cite{Linde:1993xx,Garcia-Bellido:1994ci,Vilenkin:1995yd}.
This sensitivity to the choice of the {}``time gauge'' $\tau$ is
unavoidable since hypersurfaces of equal $\tau$ may preferentially
select regions with certain properties. For instance, most of the
proper volume in equal-$t$ hypersurfaces is filled with regions that
have gained expansion by remaining near the top of the potential $V(\phi)$,
while hypersurfaces of equal scale factor will under-represent those
regions. Thus, a statement such as {}``most of the volume in the
Universe has values of $\phi$ with high $V(\phi)$'' is largely
gauge-dependent.

In the early works on eternal inflation~\cite{Linde:1993nz,Linde:1993xx,Garcia-Bellido:1994ci,Garcia-Bellido:1994vz},
the late-time asymptotic distribution of volume $P_{V}^{(\tilde{\gamma})}(\phi)$
along hypersurfaces of equal proper time {[}see Eq.~(\ref{eq:late-time V}){]}
was interpreted as the stationary distribution of field values in
the universe. However, the high sensitivity of this distribution to
the choice of the time variable makes this interpretation unsatisfactory.
Also, it was noted~\cite{Tegmark:2004qd} that equal-proper time
volume distributions predict an unacceptably small probability for
the currently observed CMB temperature. The reason for this result
is the extreme bias of the proper-time gauge towards over-representing
regions where reheating occurred very recently~\cite{Linde:1994gy,Vilenkin:1995yd}.
One might ask whether hypersurfaces of equal scale factor or some
other choice of time gauge would provide less biased answers. However,
it turns out~\cite{Winitzki:2005ya} that there exists no \emph{a
priori} choice of the time gauge $\tau$ that provides unbiased equal-$\tau$
probability distributions for all potentials $V(\phi)$ in models
of slow-roll inflation (see Sec.~\ref{sub:Predictions-in-continuous}
for details). 

Although the FP equations necessarily involve a dependence on gauge,
they do provide a useful statistical picture of the distribution of
fields in the universe. The FP techniques can also be used for deriving
several gauge-independent results. For instance, the presence of eternal
inflation is a gauge-independent statement (see also Sec.~\ref{sec:Presence-of-eternal}):
if the largest eigenvalue $\tilde{\gamma}$ is positive in one gauge
of the form~(\ref{eq:t tau}), then $\tilde{\gamma}>0$ in every
other gauge~\cite{Winitzki:2001np}. Using the FP approach, one can
also compute the fractal dimension of the inflating domain~\cite{Aryal:1987vn,Winitzki:2001np}
and the probability of exiting inflation through a particular point
$\phi_{*}$ of the reheating boundary in the configuration space (in
case there exists more than one such point). 

The exit probability can be determined as follows~\cite{Vilenkin:1999kd,Helmer:2006tz}.
Let us assume for simplicity that there are two possible exit points
$\phi_{*}$ and $\phi_{E}$, and that the initial distribution is
concentrated at $\phi=\phi_{0}$, i.e.\begin{equation}
P(\phi,t=0)=\delta(\phi-\phi_{0}),\end{equation}
where $\phi_{E}<\phi_{0}<\phi_{*}$.  The probability of exiting
inflation through $\phi=\phi_{E}$ during a time interval $[t,t+dt]$
is \begin{equation}
dp_{\text{exit}}(\phi_{E})=-v(\phi_{E})P(\phi_{E},t)dt\end{equation}
(note that $v(\phi_{E})<0$). Hence, the total probability of exiting
through $\phi=\phi_{E}$ at any time is\begin{equation}
p_{\textrm{exit}}(\phi_{E})=\int_{0}^{\infty}dp_{\text{exit}}(\phi_{E})=-v(\phi_{E})\int_{0}^{\infty}P(\phi_{E},t)dt.\end{equation}
Introducing an auxiliary function $F(\phi)$ as\begin{equation}
F(\phi)\equiv-v(\phi)\int_{0}^{\infty}P(\phi,t)dt,\end{equation}
one can show that $F(\phi)$ satisfies the gauge-invariant equation,
\begin{equation}
\partial_{\phi}\left[\partial_{\phi}\left(\frac{D}{v}F\right)-F\right]=\delta(\phi-\phi_{0}).\label{eq:for p exit}\end{equation}
This is in accord with the fact that $p_{\textrm{exit}}(\phi_{E})=F(\phi_{E})$
is a gauge-invariant quantity. Equation~(\ref{eq:for p exit}) with
the boundary conditions\begin{equation}
F(\phi_{*})=0,\quad\left.\partial_{\phi}(\frac{D}{v}F)\right|_{\phi=\phi_{E}}=0,\end{equation}
 can be straightforwardly integrated and yields explicit expressions
for the exit probability $p_{\text{exit}}(\phi_{E})$ as a function
of the initial value $\phi_{0}$~\cite{Helmer:2006tz}. The exit
probability $p_{\text{exit}}(\phi_{*})$ can be determined similarly.

\subsection{Self-reproduction of tunneling type\label{sub:Bubble-nucleation-models}}

Until now, we considered eternal self-reproduction due to random walk
of a scalar field. Another important class of models includes self-reproduction
due to bubble nucleation.%
\footnote{Both processes may be combined in a single scenario~\cite{Garriga:1997ef},
but we shall consider them separately for clarity.%
} Such scenarios of eternal inflation were studied in Refs.~\cite{Gott:1982zf,Guth:1982pn,Gott:1984ps,Bucher:1994gb,Bucher:1995et,Yamamoto:1995sw}.

In a locally de Sitter universe dominated by dark energy, nucleation
of bubbles of false vacuum may occur due to tunneling~\cite{Coleman:1977py,Coleman:1980aw,Hawking:1982my,Lee:1987qc}.
Since the bubble nucleation rate $\kappa$ per unit 4-volume is very
small~\cite{Coleman:1980aw,Garriga:1993fh},\begin{equation}
\kappa=O(1)H^{-4}\exp\left[-S_{I}-\frac{\pi}{H^{2}}\right],\label{eq:nucleation rate}\end{equation}
where $S_{I}$ is the instanton action and $H$ is the Hubble constant
of the de Sitter background, bubbles will generically not merge into
a single false-vacuum domain~\cite{Guth:1982pn}. Hence, infinitely
many bubbles will be nucleated at different places and times. The
resulting {}``daughter'' bubbles may again contain an asymptotically
de Sitter, infinite universe, which again gives rise to infinitely
many {}``grand-daughter'' bubbles. This picture of eternal self-reproduction
was called the {}``recycling universe''~\cite{Garriga:1997ef}.
Some (or all) of the created bubbles may support a period of additional
inflation followed by reheating, as shown in Fig.~\ref{cap:A-spacetime-diagram-bubble}.

\begin{figure}
\begin{center}\psfrag{L}{$\Lambda$} \psfrag{X}{$X$} \psfrag{0}{0}\includegraphics[%
  width=0.50\columnwidth]{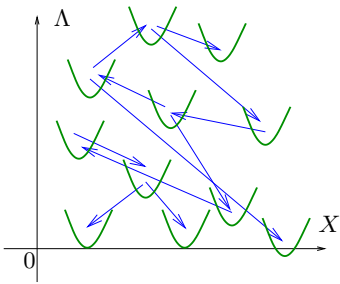}\end{center}

\caption{A schematic representation of the {}``landscape of string theory,''
consisting of a large number of local minima of an effective potential.
The variable $X$ collectively denotes various fields and $\Lambda$
is the effective cosmological constant. Arrows show possible tunneling
transitions between vacua.\label{cap:landscape}}
\end{figure}

\begin{figure}
\begin{center}\psfrag{1}{1} \psfrag{2}{2} \psfrag{3}{3} \psfrag{4}{4} \psfrag{5}{5} \psfrag{0}{0} \includegraphics[%
  width=0.80\columnwidth]{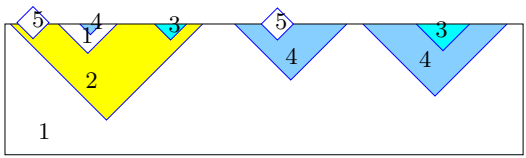}\end{center}

\caption{A conformal diagram of the spacetime where self-reproduction occurs
via bubble nucleation. Regions labeled {}``5'' are asymptotically
flat ($\Lambda=0$). \label{cap:landscape-conf}}
\end{figure}

In the model of Ref.~\cite{Garriga:1997ef}, there were only two
vacua which could tunnel into each other. A more recently developed
paradigm of {}``string theory landscape''~\cite{Susskind:2003kw}
involves a very large number of metastable vacua, corresponding to
local minima of an effective potential in field space. The value of
the potential at each minimum is the effective value of the cosmological
constant $\Lambda$ in the corresponding vacuum. Figure~\ref{cap:landscape}
shows a phenomenologist's view of the {}``landscape.'' Vacua with
$\Lambda\leq0$ do not allow any further tunneling%
\footnote{Asymptotically flat $\Lambda=0$ vacua cannot support tunneling~\cite{Farhi:1986ty,Farhi:1989yr,Bousso:2004tv};
vacua with $\Lambda<0$ will quickly collapse to a {}``big crunch''
singularity~\cite{Coleman:1980aw,Abbott:1985kr}.%
} and are called {}``terminal'' vacua~\cite{Garriga:2005av}, while
vacua with $\Lambda>0$ are called {}``recyclable'' since they can
tunnel to other vacua with $\Lambda>0$ or $\Lambda\leq0$. Bubbles
of recyclable vacua will give rise to infinitely many nested {}``daughter''
bubbles. A conformal diagram of the resulting spacetime is outlined
in Fig.~\ref{cap:landscape-conf}. Of course, only a finite number
of bubbles can be drawn; the bubbles actually form a fractal structure
in a conformal diagram~\cite{Winitzki:2005fy}.


A statistical description of the {}``recycling'' spacetime can
be obtained~\cite{Garriga:1997ef,Garriga:2005av} by considering
a single comoving worldline $\mathbf{x}=\textrm{const}$ that passes
through different bubbles at different times. (It is implied that
the worldline is randomly chosen from an ensemble of infinitely many
such worldlines passing through different points $\mathbf{x}$.) Let
the index $\alpha=1,...,N$ label all the available types of bubbles.
For calculations, it is convenient to use the $e$-folding time $\tau\equiv\ln a$.
We are interested in the probability $f_{\alpha}(\tau)$ of passing
through a bubble of type $\alpha$ at time $\tau$. This probability
distribution is normalized by $\sum_{\alpha}f_{\alpha}=1$; the quantity
$f_{\alpha}(\tau)$ can be also visualized as the fraction of the
comoving volume occupied by bubbles of type $\alpha$ at time $\tau$.
Denoting by $\kappa_{\alpha\beta}$ the nucleation rate for bubbles
of type $\alpha$ within bubbles of type $\beta$ {[}computed according
to Eq.~(\ref{eq:nucleation rate}){]}, we write the {}``master equation''
describing the evolution of $f_{\alpha}(\tau)$,\begin{equation}
\frac{df_{\beta}}{d\tau}=\sum_{\alpha}\left(-\kappa_{\alpha\beta}f_{\beta}+\kappa_{\beta\alpha}f_{\alpha}\right)\equiv\sum_{\alpha}M_{\beta\alpha}f_{\alpha},\label{eq:master equation}\end{equation}
where we introduced the auxiliary matrix $M_{\alpha\beta}$. Given
a set of initial conditions $f_{\alpha}(0)$, one can evolve $f_{\alpha}(\tau)$
according to Eq.~(\ref{eq:master equation}). 

To proceed further, one may now distinguish the following two cases:
Either terminal vacua exist (some $\beta$ such that $\kappa_{\alpha\beta}=0$
for all $\alpha$), or all the vacua are recyclable. (Theory suggests
that the former case is more probable~\cite{Kachru:2003aw}.) If
terminal vacua exist, then the late-time asymptotic solution can be
written as~\cite{Garriga:2005av} \begin{equation}
f_{\alpha}(\tau)\approx f_{\alpha}^{(0)}+s_{\alpha}e^{-q\tau},\label{eq:def s q}\end{equation}
where $f_{\alpha}^{(0)}$ is a constant vector that depends on the
initial conditions and has nonzero components only in terminal vacua,
and $s_{\alpha}$ does not depend on initial conditions and is an
eigenvector of $M_{\alpha\beta}$ such that $\sum_{\alpha}M_{\beta\alpha}s_{\alpha}=-qs_{\beta}$,
$q>0$. This solution shows that all comoving volume reaches terminal
vacua exponentially quickly. (As in the case of random-walk inflation,
there are infinitely many {}``eternally recycling'' points $\mathbf{x}$
that never enter any terminal vacua, but these points form a set of
measure zero.)

If there are no terminal vacua, the solution $f_{\alpha}(\tau)$ approaches
a constant distribution~\cite{Vanchurin:2006qp},\begin{align}
\lim_{\tau\rightarrow\infty}f_{\alpha}(\tau) & \approx f_{\alpha}^{(0)},\quad\sum_{\beta}M_{\beta\alpha}f_{\alpha}^{(0)}=0,\label{eq:fa0}\\
f_{\alpha}^{(0)} & =H_{\alpha}^{4}\exp\left[\frac{\pi}{H_{\alpha}^{2}}\right].\end{align}
In this case, the quantities $f_{\alpha}^{(0)}$ are independent of
initial conditions and are interpreted as the fractions of time spent
by the comoving worldline in bubbles of type $\alpha$.

One may adopt another approach and ignore the duration of time spent
by the worldline within each bubble. Thus, one describes only the
\emph{sequence} of the bubbles encountered along a randomly chosen
worldline~\cite{Vanchurin:2006qp,Bousso:2006ev,Aguirre:2006ak}.
If the worldline is initially in a bubble of type $\alpha$, then
the probability $\mu_{\beta\alpha}$ of entering the bubble of type
$\beta$ as the next bubble in the sequence after $\alpha$ is \begin{equation}
\mu_{\beta\alpha}=\frac{\kappa_{\beta\alpha}}{\sum_{\gamma}\kappa_{\gamma\alpha}}.\end{equation}
(For terminal vacua $\alpha$, we have $\kappa_{\gamma\alpha}=0$
and so we may define $\mu_{\beta\alpha}=0$ for convenience.) Once
again we consider landscapes without terminal vacua separately from
terminal landscapes. If there are no terminal vacua, then the matrix
$\mu_{\alpha\beta}$ is normalized, $\sum_{\beta}\mu_{\beta\alpha}=1$,
and is thus a stochastic matrix~\cite{Lancaster:1969:ToM} describing
a Markov process of choosing the next visited vacuum. The sequence
of visited vacua is infinite, so one can define the mean frequency
$f_{\alpha}^{(\text{mean})}$ of visiting bubbles of type $\alpha$.
If the probability distribution for the first element in the sequence
is $f_{(0)\alpha}$, then the distribution of vacua after $k$ steps
is given (in the matrix notation) by the vector\begin{equation}
\mathbf{f}_{(k)}=\boldsymbol{\mu}^{k}\mathbf{f}_{(0)},\end{equation}
where $\boldsymbol{\mu}^{k}$ means the $k$-th power of the matrix
$\boldsymbol{\mu}\equiv\mu_{\alpha\beta}$. Therefore, the mean frequency
of visiting a vacuum $\alpha$ is computed as an average of $f_{(k)\alpha}$
over $n$ consecutive steps in the limit of large $n$:\begin{equation}
\mathbf{f}^{(\text{mean})}=\lim_{n\rightarrow\infty}\frac{1}{n}\sum_{k=1}^{n}\mathbf{f}_{(k)}=\lim_{n\rightarrow\infty}\frac{1}{n}\sum_{k=1}^{n}\boldsymbol{\mu}^{k}\mathbf{f}_{(0)}.\label{eq:limit frequency}\end{equation}
(It is proved in the theory of Markov processes that the limit $f_{\alpha}^{(\text{mean})}$
given by Eq.~(\ref{eq:limit frequency}) almost surely coincides
with the mean frequency of visiting the state $\alpha$; see e.g.~Ref.~\cite{Doob:1953:SP},
chapter 5, Theorem 2.1, and Ref.~\cite{Kannan:1979:SP}, Theorem
3.5.9.) It turns out that the distribution $\mathbf{f}^{(\text{mean})}$
is independent of the initial state $\mathbf{f}_{(0)}$ and coincides
with the distribution~(\ref{eq:fa0}) found in the continuous-time
description~\cite{Vanchurin:2006qp}. 

If there exist terminal vacua, then almost all sequences will have
a finite length. The distribution of vacua in a randomly chosen sequence
is still well-defined and can be computed using Eq.~(\ref{eq:limit frequency})
without the normalizing factor $\frac{1}{n}$,\begin{equation}
\mathbf{f}^{(\text{mean})}=\left(\mathbf{1}-\boldsymbol{\mu}\right)^{-1}\boldsymbol{\mu}\mathbf{f}_{(0)},\label{eq:f worldline}\end{equation}
 but now the resulting distribution depends on the initial state $\mathbf{f}_{(0)}$~\cite{Bousso:2006ev,Aguirre:2006ak}.

\section{Predictions and measure issues \label{sec:Predictions-and-cutoffs}}

As discussed in Sec.~\ref{sub:Physical-relevance}, a compelling
question in the context of eternal inflation is how to make statistical
predictions of observed parameters. One begins by determining whether
eternal inflation is present in a given model.

\subsection{Presence of eternal inflation\label{sec:Presence-of-eternal}\label{sub:Global-structure-of}}

\label{eternal points}Since the presence of eternal self-reproduction
in models of tunneling type is generically certain (unless the nucleation
rate for terminal vacua is unusually high), in this section we restrict
our attention to eternal inflation of the random-walk type.

The hallmark of eternal inflation is the unbounded growth the total
number of independent inflating regions. The total proper 3-volume
of the inflating domain also grows without bound at late times, at
least when computed along hypersurfaces of equal proper time or equal
scale factor. However, the proper 3-volume is a gauge-dependent quantity,
and one may construct time gauges where the 3-volume decreases with
time even in an everywhere expanding universe~\cite{Winitzki:2005ya}.
The 3-volume of an \emph{arbitrary} family of equal-time hypersurfaces
cannot be used as a criterion for the presence of eternal inflation.
However, a weaker criterion is sufficient: Eternal inflation is present
if (and only if) there \emph{exists} a choice of time slicing with
an unbounded growth of the 3-volume of inflating domains~\cite{Winitzki:2005ya}.
Equivalently, eternal inflation is present if a finite comoving volume
gives rise to infinite physical volume~\cite{Tegmark:2004qd}. Thus,
the presence or absence of eternal inflation is a gauge-independent
statement. One may, of course, use a particular gauge (such as the
proper time or $e$-folding time) for calculations, as long as the
result is known to be gauge-independent. It can be shown that eternal
inflation is present if and only if the 3-volume grows in the $e$-folding
time slicing~\cite{Winitzki:2005ya}.

The presence of eternal inflation has been analyzed in many specific
scenarios. For instance, eternal inflation is generic in {}``chaotic''~\cite{Linde:1984st,Linde:1986fc,Linde:1986fd}
and {}``new''~\cite{Vilenkin:1983xp} inflationary models. It is
normally sufficient to establish the existence of a {}``diffusion-dominated''
regime, that is, a range of $\phi$ where the typical amplitude $\delta\phi\sim H$
of {}``jumps'' is larger than the typical change of the field, $\dot{\phi}\Delta t$,
during one Hubble time $\Delta t=H^{-1}$. For models of scalar-field
inflation, the condition is\begin{equation}
H^{2}\gg H'.\end{equation}
 Such a range of $\phi$ is present in most slow-roll models of inflation.
(For an example of an inflationary scenario where eternal inflation
is generically \emph{not} present, see Ref.~\cite{Chen:2006hs}.)
A strict formal criterion for the presence of eternal inflation is
the positivity of the largest eigenvalue $\tilde{\gamma}$ of the
operator $\hat{L}+3H(\phi)$, as defined in Sec.~\ref{sub:Methods-of-solution}.

The causal structure of the eternally inflating spacetime and the
topology of the reheating surface can be visualized using the construction
of {}``eternal comoving points''~\cite{Winitzki:2001np}. These
are comoving worldlines $\mathbf{x}=\text{const}$ that forever remain
within the inflating domain and never enter the reheating epoch. These
worldlines correspond to places where the reheating surface reaches
$t=\infty$ in a spacetime diagram (see Fig.~\ref{cap:reheating-surface-1}).
It was shown in Ref.~\cite{Winitzki:2001np} using topological arguments
that the presence of inflating domains at arbitrarily late times entails
the existence of infinitely many such {}``eternal points.'' The
set of all eternal points within a given three-dimensional spacelike
slice is a measure zero fractal set. The fractal dimension of this
set can be understood as the fractal dimension of the inflating domain~\cite{Aryal:1987vn,Winitzki:2001np,Winitzki:2005ya}
and is invariant under any \emph{smooth} coordinate transformations
in the spacetime.

The existence of eternal points can be used as another invariant criterion
for the presence of eternal inflation. The probability $X(\phi)$
of having an eternal point in an initial Hubble-size region with field
value $\phi$ can be found as the solution of a gauge-invariant, nonlinear
diffusion equation~\cite{Winitzki:2001np}\begin{equation}
D\partial_{\phi\phi}X+v\partial_{\phi}X-3H\left(1-X\right)\ln\left(1-X\right)=0,\end{equation}
with zero boundary conditions. Eternal inflation is present if there
exists a nontrivial solution $X(\phi)\not\equiv0$ of this equation.

\subsection{Observer-based measure in eternal inflation\label{sub:Predictions-of-observable}}

In theories where observable parameters $\chi_{a}$ are distributed
randomly, one would like to predict the values of $\chi_{a}$ most
likely to be observed by a random (or {}``typical'') observer. More
generally, one looks for the probability distribution $P(\chi_{a})$
of observing the values $\chi_{a}$. As discussed in Sec.~\ref{sub:Physical-relevance},
considerations of this type necessarily involve some form of the {}``principle
of mediocrity''~\cite{Vilenkin:1995yd}. On a more formal level,
one needs to construct an ensemble of the {}``possible observers''
and to define a probability measure on this ensemble. In inflationary
cosmology, observers appear only along the reheating surface. If eternal
inflation is present, the reheating surface contains infinitely many
causally disconnected and (possibly) statistically inequivalent domains.
The principal difficulties in the probabilistic approach are due to
a lack of a natural definition of measure on such surfaces.%
\footnote{To avoid confusion, let us note that the recent work~\cite{Gibbons:2006pa}
proposes a measure in the phase space of trajectories rather than
an observer-based measure in the sense discussed here.%
}

Existing proposals for an observer-based measure fall in two major
classes, which may be designated as {}``volume-based'' vs.~{}``worldline-based.''
The difference between these classes is in the approach taken to construct
the ensemble of observers. In the {}``volume'' approach~\cite{Linde:1993xx,Vilenkin:1995yd,Vilenkin:1998kr,Vanchurin:1999iv,Garriga:2005av,Easther:2005wi},
the ensemble contains every observer appearing in the universe, at
any time or place. In the {}``worldline'' approach~\cite{Garriga:2001ri,Bousso:2006ev,Bousso:2006ge,Bousso:2006nx},
the ensemble consists of observers appearing near a single, randomly
selected comoving worldline $\mathbf{x}=\text{const}$; more generally,
an arbitrary timelike geodesic could also be used. If the ensemble
contains infinitely many observers (this is typically the case for
volume-based ensembles), a regularization is needed to obtain specific
probability distributions. Finding and applying suitable regularization
procedures is a separate technical issue explored in Sec.~\ref{sub:Predictions-in-continuous}
and~\ref{sub:Predictions-in-discrete}. I begin with a general discussion
of these measure prescriptions.

A number of previously considered {}``volume-based'' measure proposals
were found to be lacking in one aspect or another~\cite{Linde:1993xx,Vilenkin:1995yd,Winitzki:1995pg,Linde:1995uf,Vilenkin:1996ar,Garriga:2001ri},
the most vexing problem being the dependence on the choice of the
time gauge~\cite{Linde:1993nz,Linde:1993xx,Tegmark:2004qd}. The
requirement of time gauge independence is sufficiently important to
reject any measure proposal that suffers from the gauge ambiguity.
A prescription manifestly free from gauge dependence is the {}``spherical
cutoff'' measure~\cite{Vilenkin:1998kr}. This prescription provides
unambiguous predictions for models of random-walk type eternal inflation
if the reheating condition $\phi=\phi_{*}$ corresponds to a topologically
compact and connected locus in the field space $\left\{ \phi,\chi_{a}\right\} $
(see Sec.~\ref{sub:Predictions-in-continuous}). For models where
the reheating condition is met at several disconnected loci in field
space (tunneling-type eternal inflation belongs to this class), one
can use the recently proposed prescription of {}``comoving cutoff''~\cite{Garriga:2005av,Easther:2005wi}.
Since no other volume-based prescriptions are currently considered
viable, we refer to the mentioned spherical cutoff/comoving cutoff
prescriptions simply as the {}``volume-based measure.''

Similarly, existing measure proposals of the {}``worldline'' type
appear to converge essentially to a single prescription~\cite{Bousso:2006ev,Aguirre:2006ak}
(however, see~\cite{Aguirre:2006na,Vanchurin:2006xp} for the most
recent developments). We refer to this prescription as the {}``worldline-based
measure.''

The main difference between the worldline-based and volume-based measures
is in their dependence on the initial state. When considering the
volume-based measure, one starts from a finite initial spacelike 3-volume.
(Final results are insensitive to the choice of this 3-surface in
spacetime or to its geometry.) The initial state consists of the initial
values of the fields $\chi_{a}$ within the initial volume, and possibly
a label $\alpha$ corresponding to the type of the initial bubble.
When considering the worldline-based measure, one assumes knowledge
of these data at the initial point of the worldline.%
\footnote{Naturally, it is assumed that the initial state is in the self-reproduction
regime: For random-walk type models, the initial 3-volume is undergoing
inflation rather than reheating and, for tunneling models, the initial
3-volume is not situated within a terminal bubble.%
} It turns out that the volume-based measure always yields results
that are independent of the initial conditions. This agrees with the
concept of the {}``stationarity'' of a self-reproducing universe~\cite{Linde:1993nz,Linde:1993xx,Garcia-Bellido:1994ci};
the universe forgets the initial state in the course of eternal self-reproduction.
In contrast, probabilities obtained using the worldline-based measure
always depend on the initial state (except for the case of a {}``nonterminal''
landscape, i.e.~a landscape scenario without terminal vacua). A theory
of initial conditions is necessary to obtain a specific prediction
from the worldline-based measure.

At this time, there is no consensus as to which of the two measures
is the physically relevant one. The present author is inclined to
regard the two measures as reasonable answers to two differently posed
questions. The first question is to determine the probability distribution
for observed values of $\chi_{a}$, given that the observer is randomly
chosen from all the observers present in the entire spacetime. Since
we have no knowledge as to the total duration of inflation in our
past or the total number of bubble nucleations preceding the most
recent one, it appears reasonable to include in the ensemble all the
observers that will ever appear anywhere in the spacetime. The answer
to the first question is thus provided by the volume-based measure. 

The second question is posed in a rather different manner. In the
context of tunneling-type eternal inflation, upon discovering the
type of our bubble we may wish to leave a message to a future civilization
that may arise in our future after an unspecified number of nested
bubble nucleations. The analogous situation in the context of random-walk
type inflation is a hypothetical observer located within an inflating
region of spacetime who wishes to communicate with future civilizations
that will eventually appear when inflation is over. The only available
means of communication is leaving information on paper in a sealed
box. The message might contain the probability distribution $P(\chi_{a})$
for parameters $\chi_{a}$ that we expect the future civilization
to observe. In this case, the initial state is known at the time of
writing the message. It is clear that the box can be discovered only
by future observers near its worldline. It is then natural to choose
the ensemble of observers appearing along this worldline. Starting
from the known initial state, one would then compute $P(\chi_{a})$
according to the worldline-based measure.

Calculations using the worldline-based measure usually do not require
regularization (except for the case of nonterminal landscape) because
the worldline-based ensemble of observers is almost surely finite~\cite{Bousso:2006ev}.
For instance, in random-walk type models the worldline-based measure
predicts the exit probability distribution $p_{\text{exit}}$, which
can be computed by solving a suitable differential equation {[}see
Eq.~(\ref{eq:for p exit}){]}. However, the ensemble used in the
volume-based measure is infinite and requires a regularization. Known
regularization methods are reviewed in Sections~\ref{sub:Predictions-in-continuous}
and~\ref{sub:Predictions-in-discrete}.

\begin{figure}
\begin{center}\psfrag{phi}{$\phi$}\psfrag{V}{$V$}

\psfrag{ps2}{$\phi_{*}^{(1)}$}\psfrag{ps1}{$\phi_{*}^{(2)}$}

\psfrag{p2}{$\phi_1$}\psfrag{p1}{$\phi_2$}\includegraphics[%
  width=0.70\columnwidth]{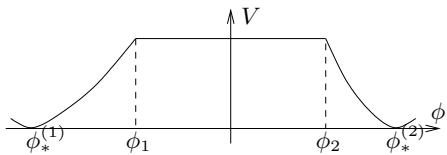}\end{center}

\caption{Illustrative inflationary potential with a flat self-reproduction
regime $\phi_{1}<\phi<\phi_{2}$ and deterministic regimes $\phi_{*}^{(1)}<\phi<\phi_{1}$
and $\phi_{2}<\phi<\phi_{*}^{(2)}$. \label{cap:pot1}}
\end{figure}

A simple toy model~\cite{Vilenkin:1995yd} where the predictions
of the volume-based measure can be obtained analytically is a slow-roll
scenario with a potential shown in Fig.~\ref{cap:pot1}. The potential
is flat in the range $\phi_{1}<\phi<\phi_{2}$ where the evolution
is diffusion-dominated, while the evolution of regions with $\phi>\phi_{2}$
or $\phi<\phi_{1}$ is completely deterministic (fluctuation-free).
It is assumed that the diffusion-dominated range $\phi_{1}<\phi<\phi_{2}$
is sufficiently wide to cause eternal self-reproduction of inflating
regions. There are two thermalization points, $\phi=\phi_{*}^{(1)}$
and $\phi=\phi_{*}^{(2)}$, which may be associated to different types
of true vacuum and thus to different observed values of cosmological
parameters. The question is to compare the volumes ${\cal V}_{1}$
and ${\cal V}_{2}$ of regions thermalized into these two vacua. Since
there is an infinite volume thermalized into either vacuum, one looks
for the volume ratio ${\cal V}_{1}/{\cal V}_{2}$.

The potential is symmetric in the range $\phi_{1}<\phi<\phi_{2}$,
so it is natural to assume that Hubble-size regions exiting the self-reproduction
regime at $\phi=\phi_{1}$ and at $\phi=\phi_{2}$ are equally abundant.
Since the evolution within the ranges $\phi_{*}^{(1)}<\phi<\phi_{1}$
and $\phi_{2}<\phi<\phi_{*}^{(2)}$ is deterministic, the regions
exiting the self-reproduction regime at $\phi=\phi_{1}$ or $\phi=\phi_{2}$
will be expanded by fixed amounts of $e$-foldings, which we may denote
$N_{1}$ and $N_{2}$ respectively, \begin{equation}
N_{j}=-4\pi\int_{\phi_{j}}^{\phi_{*}^{(j)}}\frac{H(\phi)}{H'(\phi)}d\phi.\end{equation}
 Therefore the volume of regions thermalized at $\phi=\phi_{*}^{(j)}$,
where $j=1,2$, will be increased by the factors $\exp(3N_{j})$.
Hence, the volume ratio is\begin{equation}
\frac{{\cal V}_{1}}{{\cal V}_{2}}=\exp\left(3N_{1}-3N_{2}\right).\label{eq:ans2}\end{equation}

\subsection{Regularization for a single reheating surface\label{sub:Predictions-in-continuous}}

The task at hand is to define a measure that ascribes equal weight
to each observer ever appearing anywhere in the universe. As discussed
in Sec.~\ref{sub:Physical-relevance}, it is sufficient to construct
a measure $\mathcal{V}(\chi_{a})$ of the 3-volume along the reheating
surface. The volume-based measure $P(\chi_{a})$ will then be given
by Eq.~(\ref{eq:probability}). In the presence of eternal inflation,
the proper 3-volume of the reheating surface diverges even when we
limit the spacetime domain under consideration to the comoving future
of a finite initial spacelike 3-volume. Therefore, the reheating surface
needs to be regularized.

\begin{figure}
\begin{center}\includegraphics[%
  width=0.60\columnwidth]{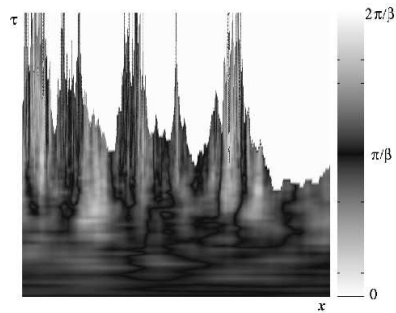}\end{center}

\caption{A 1+1-dimensional slice of spacetime in a two-field inflationary
model (numerical simulation in Ref.~\cite{Vanchurin:1999iv}). Shades
denote different values of the field $\chi$, which takes values in
the periodically identified interval $\left[0,2\pi/\beta\right]$.
The white region represents the thermalized domain. The boundary of
the thermalized domain is the reheating surface (cf.~Fig.~\ref{cap:reheating-surface-1}),
which contains all the possible values of the field $\chi$.\label{cap:reheating-surface-2}}
\end{figure}

\begin{figure}
\begin{center}\psfrag{c}{$\chi$}\psfrag{f}{$\phi$}\psfrag{f=f*}{$\phi=\phi_{*}$}\includegraphics[%
  width=0.40\columnwidth]{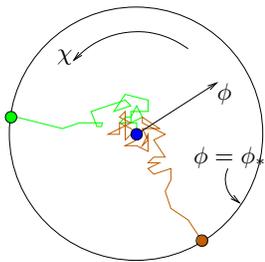}\end{center}

\caption{A random walk in configuration space for a two-field inflationary
model considered in Ref.~\cite{Vanchurin:1999iv}. The center of
the field space is a diffusion-dominated regime. The reheating condition,
$\phi=\phi_{*}$, selects a compact and connected region (a circle)
in configuration space. The problem is to determine the volume-weighted
probability distribution for the values of $\chi$ at reheating.\label{cap:random-walk}}
\end{figure}

In this section we consider the case when the reheating condition
is met at a topologically compact and connected locus in the configuration
space $\left\{ \phi,\chi_{a}\right\} $. In this case, every connected
component of the reheating surface in spacetime will contain all the
possible values of the fields $\chi_{a}$, and all such connected
pieces are statistically equivalent. Hence, it suffices to consider
a single connected piece of the reheating surface. A situation of
this type is illustrated in Fig.~\ref{cap:reheating-surface-2}.
A sketch of the random walk in configuration space is shown in Fig.~\ref{cap:random-walk}.


A simple regularization scheme is known as the {}``equal-time cutoff.''
One considers the part of the reheating surface formed before a fixed
time $t_{\max}$; that part is finite as long as $t_{\max}$ is finite.
Then one can compute the distribution of the quantities of interest
within that part of the reheating surface. Subsequently, one takes
the limit $t_{\max}\rightarrow\infty$. The resulting distribution
can be found from the solution $P_{V}^{(\tilde{\gamma})}(\phi)$ of
the {}``stationary'' FP equation, \begin{equation}
[\hat{L}+3H(\phi)]P_{V}^{(\tilde{\gamma})}=\tilde{\gamma}P_{V}^{(\tilde{\gamma})},\end{equation}
with the largest eigenvalue $\tilde{\gamma}$ (see Sec.~\ref{sub:Methods-of-solution}).
However, both the eigenvalue $\tilde{\gamma}$ and the distribution
$P_{V}^{(\tilde{\gamma})}(\phi)$ depend rather sensitively on the
choice of the equal-time hypersurfaces. Since there appears to be
no preferred choice of the cutoff hypersurfaces in spacetime, the
equal-time cutoff cannot serve as an unbiased measure. Also, it was
shown in Ref.~\cite{Winitzki:2005ya} that the unbiased result~(\ref{eq:ans2})
cannot be obtained via an equal-time cutoff with any choice of the
time gauge. 

The {}``spherical cutoff'' measure prescription~\cite{Vilenkin:1998kr}
regularizes the reheating surface in a different way. A finite region
within the reheating surface is selected as a spherical region of
radius $R$ around a randomly chosen center point. (Since the reheating
3-surface is spacelike, the distance between points can be calculated
as the length of the shortest path within the reheating 3-surface.)
Then the distribution of the quantities of interest is computed within
the spherical region. Subsequently, the limit $R\rightarrow\infty$
is evaluated. Since every portion of the reheating surface is statistically
the same, the results are independent of the choice of the center
point. The spherical cutoff is gauge-invariant since it is formulated
entirely in terms of the intrinsic properties of the reheating surface.

While the spherical cutoff prescription successfully solves the problem
of regularization, there is no universally applicable analytic formula
for the resulting distribution. Application of the spherical cutoff
to general models of random-walk inflation requires a direct numerical
simulation of the stochastic field dynamics in the inflationary spacetime.
Such simulations were reported in Refs.~\cite{Aryal:1987vn,Linde:1993xx,Linde:1994wt,Vanchurin:1999iv}
and used the Langevin equation~(\ref{eq:Langevin}) with a specific
stochastic ansatz for the noise field $N(t,\mathbf{x})$.

Apart from numerical simulations, the results of the spherical cutoff
method may be obtained analytically in a certain class of models~\cite{Vanchurin:1999iv}.
One such case is a multi-field model where the potential $V(\phi,\chi_{a})$
is independent of $\chi_{a}$ within the range of $\phi$ where the
{}``diffusion'' in $\phi$ dominates. Then the distribution in $\chi_{a}$
is flat when field $\phi$ exits the regime of self-reproduction and
resumes the deterministic slow-roll evolution. One can derive a gauge-invariant
Fokker-Planck equation for the volume distribution $P_{V}(\phi,\chi_{a})$,
using $\phi$ as the time variable~\cite{Vanchurin:1999iv},\begin{equation}
\partial_{\phi}P_{V}=\partial_{\chi\chi}\left(\frac{D}{v_{\phi}}P_{V}\right)-\partial_{\chi}\left(\frac{v_{\chi}}{v_{\phi}}P_{V}\right)+\frac{3H}{v_{\phi}}P_{V},\label{eq:FP equation phi}\end{equation}
where $D$, $v_{\phi}$, and $v_{\chi}$ are the coefficients of the
FP equation~(\ref{eq:FP 2 field}). This equation is valid for the
range of $\phi$ where the evolution of $\phi$ is free of fluctuations.
By solving Eq.~(\ref{eq:FP equation phi}), one can calculate the
volume-based distribution of $\chi_{a}$ predicted by the spherical
cutoff method as $P_{V}(\phi_{*},\chi)$. Note that the mentioned
restriction on the potential $V(\phi,\chi_{a})$ is important. In
general, the field $\phi$ cannot be used as the time variable since
the surfaces of constant $\phi$ are not everywhere spacelike due
to large fluctuations of $\phi$ in the diffusion-dominated regime.

\subsection{Regularization for multiple types of reheating surfaces\label{sub:Predictions-in-discrete}}

Let us begin by considering a simpler example: an inflationary scenario
with an \emph{asymmetric} slow-roll potential $V(\phi)$ having two
minima $\phi_{*}^{(1)}$, $\phi_{*}^{(2)}$. This scenario has two
possibilities for thermalization, possibly differing in the observable
parameters $\chi_{a}$. More generally, one may consider a scenario
with $n$ different minima of the potential, possibly representing
$n$ distinct reheating scenarios. It is important that the minima
$\phi_{*}^{(j)}$, $j=1,...,n$ are topologically disconnected \emph{in
the configuration space}. This precludes the possibility that different
minima are reached within one connected component of the reheating
hypersurface in spacetime. Additionally, the fields $\chi_{a}$ may
fluctuate across each connected component in a way that depends on
the minimum $j$. Thus, the distribution $P(\chi_{a};j)$ of the fields
$\chi_{a}$ at each connected component of the reheating surface may
depend on $j$. In other words, the different components of the reheating
surface may be statistically inequivalent with respect to the distribution
of $\chi_{a}$ on them. To use the volume-based measure for making
predictions in such models, one needs a regularization method that
is applicable to situations with a large number of disconnected and
statistically inequivalent components of the reheating hypersurface.

In such situations, the spherical cutoff prescription (see Sec.~\ref{sub:Predictions-in-continuous})
yields only the distribution of $\chi_{a}$ across one connected component,
since the sphere of a finite radius $R$ will never reach any other
components of the reheating surface. Therefore, the spherical cutoff
needs to be supplemented by a {}``weighting'' prescription, which
would assign a weights $p_{j}$ to the minimum labeled $j$. In scenarios
of tunneling type, such as the string theory landscape, observers
may find themselves in bubbles of types $\alpha=1,...,N$. Again,
a weighting prescription is needed to determine the probabilities
$p_{\alpha}$ of being in a bubble of type $\alpha$. 

Two different weighting prescriptions have been formulated, using
the volume-based~\cite{Garriga:2005av,Easther:2005wi} and the worldline-based
approach~\cite{Bousso:2006ev}, respectively. The first prescription
is called the {}``comoving cutoff'' while the second the {}``worldline''
or the {}``holographic'' cutoff. For clarity, we illustrate these
weighting prescriptions on models of tunneling type where infinitely
many nested bubbles of types $\alpha=1,...N$ are created and where
some bubbles are {}``terminal,'' i.e.~contain no {}``daughter''
bubbles. In the volume-based approach, each bubble receives equal
weight in the ensemble; in the worldline-based approach, only bubbles
intersected by a selected worldline are counted and given equal weight.
Let us now examine these two prescriptions in more detail.

\begin{figure}
\begin{center}\psfrag{t}{$t$} \psfrag{x}{$x$} \psfrag{0}{0}\includegraphics[%
  width=0.60\columnwidth]{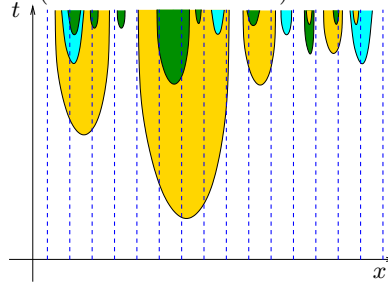}\end{center}

\caption{Weighting prescriptions for models of tunneling type. Shaded regions
are bubbles of different types. Dashed vertical lines represent randomly
chosen comoving geodesics used to define a finite subset of bubbles.
\label{cap:bubbles2}}
\end{figure}

Since the set of all bubbles is infinite, one needs to perform a {}``regularization,''
that is, one needs to select a very large but finite subset of bubbles.
The weight $p_{\alpha}$ will be calculated as the fraction of bubbles
of type $\alpha$ within the selected subset; then the number of bubbles
in the subset will be taken to infinity. Technically, the two prescriptions
differ in the details of the regularization. The difference between
the two prescriptions can be understood pictorially (Fig.~\ref{cap:bubbles2}).
In a spacetime diagram, one draws a finite number of timelike comoving
geodesic worldlines emitted from an initial 3-surface towards the
future. (It can be shown that the results are independent of the choice
of these lines, as long as that choice is uncorrelated with the bubble
nucleation process~\cite{Easther:2005wi}.) Each of these lines will
intersect only a finite number of bubbles, since the final state of
any worldline is (almost surely) a terminal bubble. The subset of
bubbles needed for the regularization procedure is defined as the
set of all bubbles intersected by at least one line. At this point,
the volume-based approach assigns equal weight to each bubble in the
subset, while the worldline-based approach assigns equal weight to
each bubble along each worldline. As a result, the volume-based measure
counts each bubble in the subset only once, while the worldline-based
measure counts each bubble as many times as it is intersected by some
worldlines. After determining the weights $p_{\alpha}$ by counting
the bubbles as described, one increases the number of worldlines to
infinity and evaluates the limit values of $p_{\alpha}$.

It is clear that the volume-based measure represents the counting
of bubbles in the entire universe, and it is then appropriate that
each bubble is being counted only once. On the other hand, the worldline-based
measure counts bubbles occurring along a single worldline, ignoring
the bubbles produced in other parts of the universe and introducing
an unavoidable bias due to the initial conditions at the starting
point of the worldline.

Explicit formulas for $p_{\alpha}$ were derived for a tunneling-type
scenario (with terminal vacua) in the volume-based approach~\cite{Garriga:2005av},\begin{equation}
p_{\alpha}=\sum_{\beta}H_{\beta}^{q}\kappa_{\alpha\beta}s_{\beta},\end{equation}
where $\kappa_{\alpha\beta}$ is the matrix of nucleation rates, $q$
and $s_{\alpha}$ are the quantities defined by Eq.~(\ref{eq:def s q}),
and $H_{\beta}$ is the Hubble parameter in bubbles of type $\beta$.
The expressions for $p_{\alpha}$ obtained from the worldline-based
measure are given by Eq.~(\ref{eq:f worldline}). As we have noted
before, the volume-based measure assigns weights $p_{\alpha}$ that
are independent of initial conditions, while the weights obtained
from the worldline-based measure depend sensitively on the type of
bubble where the counting begins.

In the case of a non-terminal landscape, both the volume-based and
the world-line based measures give identical results for $p_{\alpha}$,
which coincide with the mean frequency~(\ref{eq:fa0}) of visiting
a bubble of type $\alpha$~\cite{Vanchurin:2006qp,Bousso:2006ev}.

With the weighting prescriptions just described, the volume-based
and the worldline-based measure proposals can be considered complete.
In other words, we have two alternative prescriptions that can be
applied (in principle) to arbitrary models of random-walk or tunneling-type
eternal inflation. Further research is needed to reach a definite
conclusion concerning the viability of these measure prescriptions.

\section*{Acknowledgments\label{sec:Acknowledgments}}

The author is grateful to Gabriel Lopes Cardoso, Matthew Johnson,
and Andrei Linde for useful discussions, and Andrei Barvinsky and
Alex Vilenkin for comments on the manuscript.

\bibliographystyle{apsrev}
\bibliography{EI1}

\end{document}